\documentclass[aps,prl,amsmath,preprint,superscriptaddress]{revtex4-1} 
\usepackage{hyperref}
\usepackage{graphicx} 
\usepackage{units}
\usepackage{upgreek}
\usepackage{color}
\usepackage{xcolor}
\usepackage{amsmath,amssymb,amsfonts}
\usepackage{bbold}
\usepackage{textgreek}
\usepackage{soul}
\newcommand{\sj}[1]{{\color{black}#1}}

\newcommand{\Gammainc}{\text{\textgamma}}
\newcommand{\gammat}{\xi/M}
\newcommand{\gammar}{\xi_{\rm r}/J}
\newcommand{\delaynum}{\mathfrak{D}_0}
\newcommand{\delaynumw}{\mathfrak{D}_1}
\newcommand{\delaynumx}{\mathfrak{D}_2}
\newcommand{\difft}{D}

\begin{document}

\title{Inertial delay of self-propelled particles}

\author{Christian Scholz}
\affiliation{Institut f\"ur Theoretische Physik II: Weiche Materie, Heinrich-Heine-Universit\"at D\"usseldorf, D-40225 D\"usseldorf, Germany}

\author{Soudeh Jahanshahi}
\affiliation{Institut f\"ur Theoretische Physik II: Weiche Materie, Heinrich-Heine-Universit\"at D\"usseldorf, D-40225 D\"usseldorf, Germany}

\author{Anton Ldov}
\affiliation{Institut f\"ur Theoretische Physik II: Weiche Materie, Heinrich-Heine-Universit\"at D\"usseldorf, D-40225 D\"usseldorf, Germany}

\author{Hartmut L{\"o}wen}
\affiliation{Institut f\"ur Theoretische Physik II: Weiche Materie, Heinrich-Heine-Universit\"at D\"usseldorf, D-40225 D\"usseldorf, Germany}

\date{\today}

\begin{abstract}
The motion of self-propelled massive particles through a gaseous medium is\- do\-minated by inertial effects. Examples include vibrated granulates, activated complex plasmas and flying insects. However, inertia is usually neglected in standard models. Here, we experimentally demonstrate the significance of inertia on macroscopic self-propelled particles. We observe a distinct \textit{inertial delay} between orientation and velocity of particles, originating from the finite relaxation times in the system. This effect is fully explained by an underdamped generalisation of the Langevin model of active Brownian motion. In stark contrast to passive systems, the inertial delay profoundly influences the long-time dynamics and enables new fundamental strategies for controlling self-propulsion in active matter. 
\end{abstract}

\maketitle

Newton's first law states that because of inertia, a massive object resists any change of momentum. Before this groundbreaking idea, the dominant theory of motion was based on Aristotelian physics, which posits that objects come to rest unless propelled by a driving force. In retrospect, this perception is unsurprising, as the motions of everyday objects are influenced significantly by friction. In microscopic systems such as colloids, inertial forces are completely overwhelmed by viscous friction. In fact, biological organisms such as bacteria must self-propel by implementing non-reciprocal motion\cite{purcell1977}.\\
However, any finitely massive object performs ballistic motion, even if only on minuscule time and length scales. For example, colloidal particles undertake ballistic motion below 1\AA\ for approximately 100 ns. Experimental veri\-fication of this motion requires high accuracy measurements and has been achieved only for passive colloids\cite{Blum2006measurement,li2010measurement,huang2011direct}. However, the inertial forces of macroscopic self-propelled particles, such as animals and robots, can be comparable with propulsion forces, leading to overlap of the inertial and active motion.\\
A particularly simple example of a macroscopic self-propelled particle is a minimalistic robot called a \textit{vibrobot}, which converts vibrational energy into directed motion using its tilted elastic legs\cite{Giomi2013}. Collectives of such particles exhibit novel non-equilibrium dynamics\cite{tsai2005,Briand2016,Scholz2017,Junot2017},  self-organisation\cite{scholz2018rotating}, clustering\cite{Giomi2013,Deblais2018} and swarming\cite{kudrolli2008,deseigne2010collective,Patterson2017}. Along with artificial and biological microswimmers\cite{Cates2015,menzel2015tuned,RevModPhys.88.045006}, vibrobots belong to the class of active soft matter.\\
Here, we demonstrate that the inertia of self-propelled particles causes a significant delay between their orientation and velocity and increases the long-time diffusion coefficient through persistent correlations in the underdamped rotational motion. Standard models, such as the Viscek-model\cite{vicsek1995} and active Brownian motion\cite{golestanian2007} cannot explain this behaviour as they neglect inertia. Instead, the dynamics can be understood in terms of underdamped Langevin equations with a self-propulsion term that couples the rotational and translational degrees of freedom. Using the mean squared displacements (MSDs) and velocity distributions, fitted by numerical and analytical results, we extract a unique set of parameters for the model. We derive analytic solutions for the short- and long-time behaviour of the MSD and prove that the long-time diffusion coefficient explicitly depends on the moment of inertia.

\section{Results}

\subsection{Experimental observation of inertial effects}

\begin{figure*}[tb]
	\includegraphics[width=\textwidth]{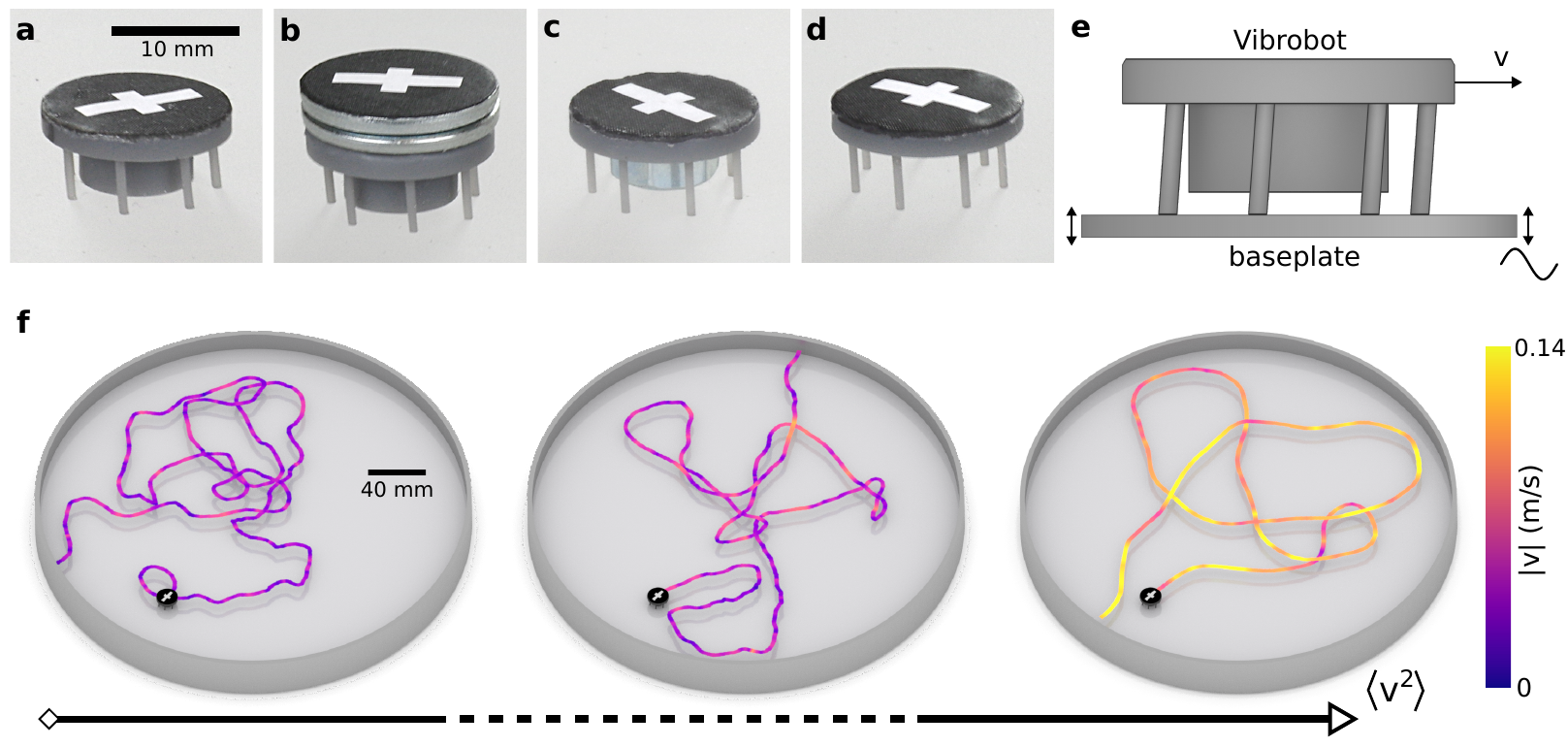}
	\caption{\label{Fig:Experiment} \textsf{\textbf{3D printed particles, setup and trajectories.} 
	\textbf{a} \textit{Generic particle}. \textbf{b} \textit{Carrier particle} with an additional outer mass. \textbf{c} \textit{Tug particle} with an additional central mass. \textbf{d} \textit{Ring particle} without a central core. \textbf{e} Illustration of the mechanism with a \textit{generic particle} on a vibrating plate. \textbf{f} Three exemplary trajectories with increasing average particle velocities. Particle images mark the starting point of each trajectory. The trajectory colour indicates the magnitude of the velocity.}}
\end{figure*}

Our experimental particles are 3D-printed vibrobots driven by sinusoidal vibrations from an electromagnetic shaker. To investigate a wide range of parameter combinations, we varied the leg inclination, mass and moment of inertia of the particles (see Fig.~\ref{Fig:Experiment}\textbf{a}-\textbf{d}). The excitation frequency and amplitude were fixed to $f=80\,\mathrm{Hz}$ and $A=66\,\mu\text{m}$, respectively, which ensures stable quasi-twodimensional motion of the particles.\\
\begin{figure*}[tb]
	\includegraphics[width=\columnwidth]{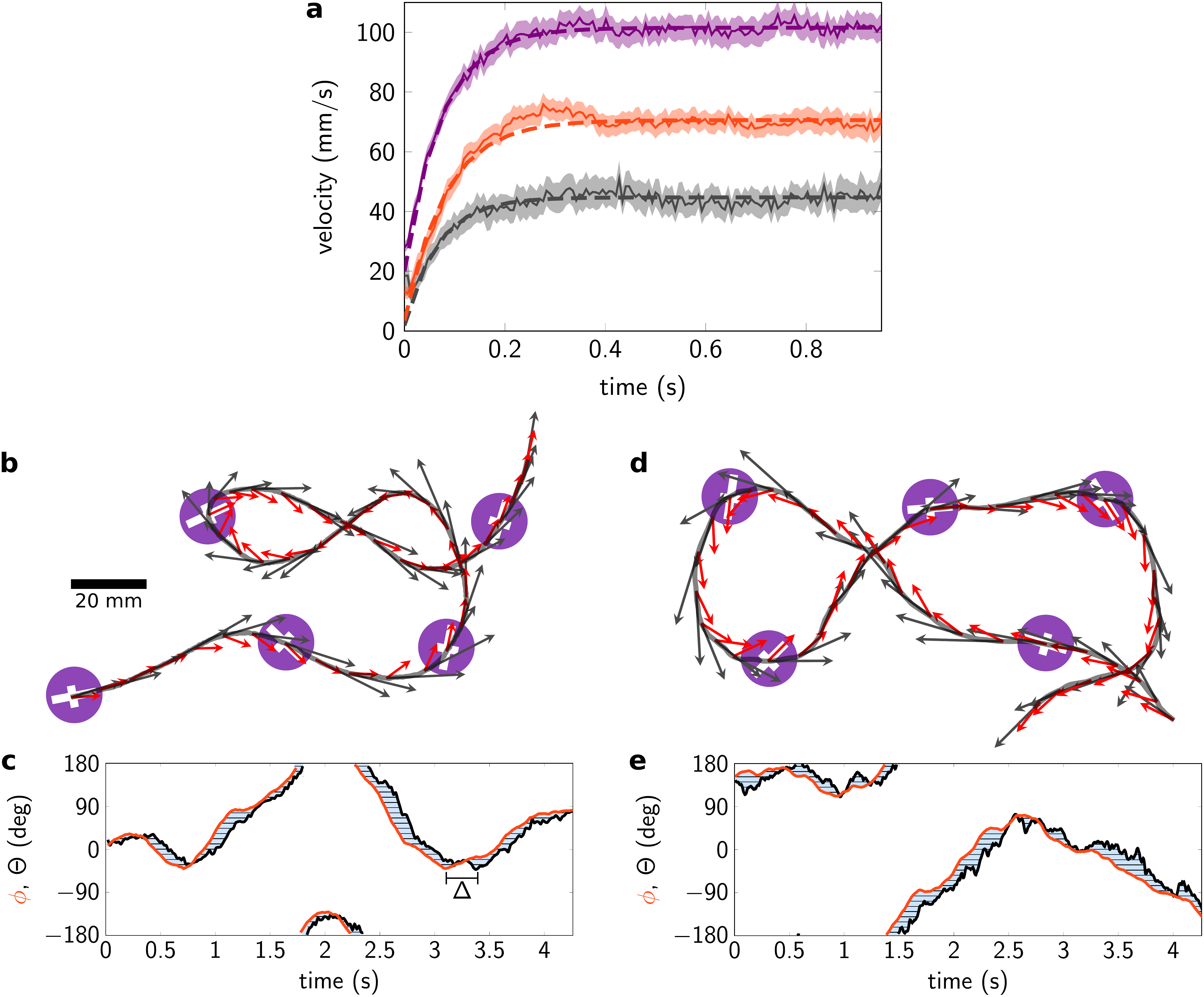}
	\caption{\label{Fig:DelayT} \textsf{\textbf{\textit{Inertial delay} in particle trajectories.} \textbf{a} Time-dependence of the average particle velocity starting from rest at $t_0$ for three particles with different leg inclinations $2$ (gray), $4$ (red) and $6$ (violet) degrees. \textbf{b} Measured particle trajectory showing the direction (black arrows) and orientation (red arrows) of a particle. \textbf{c} The measured orientation curve $\mathsf{\phi(t)}$ (red) lags the velocity direction curve $\mathsf{\Theta(t)}$ (black) by an inertial delay $\mathsf{\Delta}$. \textbf{d} Corresponding simulated trajectory with velocity direction(black arrows) and orientation(red arrows). The model parameters are $\mathsf{\gammat=6.46\,\textsf{s}^{-1}}$, $\mathsf{\gammar=5.4\,\textsf{s}^{-1}}$, $\mathsf{\difft=8\times10^{-5}\,\textsf{M}^2/\textsf{s}}$, $\mathsf{D_{\rm r}=2.59\,\textsf{s}^{-1}}$, $\mathsf{V_p=0.092\,\textsf{M}/\textsf{s}}$, $\mathsf{\omega=0.7\,\textsf{s}^{-1}}$. \textbf{e} Simulated orientation(red) and velocity curves(black).}}
\end{figure*}
The mechanism is illustrated in Fig.~\ref{Fig:Experiment}\textbf{e} and Supplementary Movie 1. The vibrobots move by a ratcheting mechanism driven by repeated collisions of their tilted elastic legs on the vibrating surface. Their propulsion velocity depends on the excitation frequency, amplitude, leg inclination and material properties such as the elasticity and friction coefficients\cite{Giomi2013,altshuler2013,koumakis2016mechanism,scholz2016ratcheting}. Long-time random motions are induced by microscopic surface inhomogeneities and (under sufficiently strong driving) a bouncing ball instability\cite{scholz2016ratcheting}. Thereby, the vibrobot motion is considered as a macroscopic realization of active Brownian motion\cite{kudrolli2008,deseigne2010collective,walsh2017noise,lanoiselee2017statistical}. Figure~\ref{Fig:Experiment}\textbf{f} shows three representative trajectories of particles with different average propulsion velocities (see also Supplementary Movie 2). The persistence length is noticeably shorter for slower particles than for faster particles, as generally expected for self-propelled particles\cite{RevModPhys.88.045006}.\\
However, the significance of inertial forces is an important difference between motile granulates and microswimmers\cite{Lam2015,Deblais2018}. Massive particles do not move instantaneously, but accelerate from rest when the vibration is started. The time-dependence of the initial velocity (averaged over up to 165 runs per particle) is shown in Fig.~\ref{Fig:DelayT}\textbf{a}. The particles noticeably accelerated up to the steady state on a time scale of $10^{-1}\,\text{s}$, one order of magnitude larger than the inverse excitation frequency and the relaxation-time of the shaker. When perturbed by an external force, vibrationally driven particles approach their steady state on a similar time scale\cite{scholz2018rotating}. The relaxation process is well fitted by an exponential function, as expected for inertial relaxation.
Inertia also influences the dynamical behaviour of the particles' orientation relative to their velocity. The orientation (red arrows in Fig.~\ref{Fig:DelayT}\textbf{b}) systematically deviates from the movement direction (black arrows in Fig.~\ref{Fig:DelayT}\textbf{b}). Particularly, during sharp turns the orientation deviates towards the centre of the curve, whereas the velocity is obviously tangential to the trajectory. We compare the angle of orientation $\phi$ to the angle of velocity $\Theta=\text{atan2}(\dot{y},\dot{x})$ in Fig.~\ref{Fig:DelayT}\textbf{b} and find that $\Theta$ systematically pursues $\phi$ with a delay of order $10^{-1}\,\text{s}$.
A slow-motion recording of one particle in Supplementary Movie 3 illustrates the dynamic delay between motion and orientation. The particle quickly reorients, but its previous direction is retained by inertia. Consequently, the particle drifts around the corner, mimicking  the well-known intentional oversteering of racing cars.\\

\subsection{Underdamped Langevin model}

Despite the complex non-linear dynamics of the vibrobots\cite{Giomi2013,Harol2016,scholz2016ratcheting,koumakis2016mechanism}, our observations can be fully described by a generalised active Brownian motion model with explicit inertial forces. The dynamics are characterized by the centre-of-mass position \sj{${\bf R}(t)=\big(X(t),Y(t)\big)$ and the orientation ${\bf n}(t)=\big(\cos \phi(t)\, ,\,\sin \phi(t)\big)$, where $\phi(t)$ defines the direction of the propulsion force.} The coupled equations of motion for ${\bf R}(t)$ and $\phi (t)$, describing the force balance between the inertial, viscous and random forces, are given by\sj{
\begin{eqnarray}
M \ddot{\bf R}(t) + \xi \dot{\bf R}(t)&=& \xi V_p {\bf n}(t) +\xi\sqrt{2D}\, {\bf f}_{\rm st}(t),\label{eq:lgm1}\\
J \ddot\phi (t) + \xi_{\rm r} \dot\phi (t)&=& \tau_0 +\xi_{\rm r}\sqrt{2D_{\rm r}}\, \tau_{\rm st}(t).\label{eq:lgm2}
\end{eqnarray}
Here, $M$ and $J$ are the mass and moment of inertia, respectively, and $\xi$ and $\xi_{\rm r}$ denote the translational and rotational friction coefficients. The translational and rotational Brownian fluctuations are quantified by their respective short--time diffusion coefficients $D$ and $D_{\rm r}$. The random forces ${\bf f}_{\rm st}(t)$ and torque $\tau_{\rm st}(t)$ are white noise terms with zero mean and correlation functions \sj{$\langle{\bf f}_{\rm st}(t)\otimes{\bf f}_{\rm st}(t')\rangle=\delta(t-t')\mathbb{1}$ and $\langle\tau_{\rm st}(t)\tau_{\rm st}(t')\rangle=\delta(t-t')$, respectively, where $\langle\, \dotsb \rangle$ denotes the ensemble average and $\mathbb{1}$ is the unit matrix. Owing to the strong non-equilibrium nature of the system, the diffusion and damping constants are not related by the Stokes-Einstein relation\cite{Jaeger1996}.} Moreover, as typical particles are not perfectly symmetrical, they tend to perform circular motions on intermediate time scales. To capture this behaviour, we applied an external torque $\tau_0$ that induces circular movement with average velocity $\omega=\tau_0/\xi_r$\cite{kummel2013circular,Kurzthaler2017}.} Similar models applied in the literature, have typically neglected the moment of inertia or have only been solved numerically \cite{weber2013long,Mokhtari2017,zhu2018transport,prathyusha2018dynamically,Parisi2018,Deblais2018}. The motion of a particle governed by Eqs.~\eqref{eq:lgm1} and \eqref{eq:lgm2} is determined by different time scales given by the friction rates $\gammat=\tau^{-1}$ and $\gammar=\tau_{\rm r}^{-1}$, the rotational diffusion rate $D_{\rm r}$, the angular frequency $\omega$ and the crossover times ${2D}/{V_p^2}$ and ${2D_{\rm r}}/{\tau_0^2}$. In the limit of vanishing $M$ and $J$ the model is equivalent to the well known active Brownian motion formulation\cite{golestanian2007}.\\
The trajectories obtained by numerically integrating the Langevin model compare well with the experimental observations. As show by the representative trajectory in Fig.~\ref{Fig:DelayT}\textbf{d},\textbf{e}, the model reproduces the delay between the orientation and velocity, when the friction is sufficiently weaker than the inertia. 
The model can be analytically solved by averaging and integration. The orientational correlation 
\begin{align}
\left\langle \mathbf{n}(t)\cdot\mathbf{n}(0)\right\rangle_T = \cos(\omega t)\,e^{\left(-D_{\rm r}(t-\tau_{\rm r}(1-e^{-t/\tau_{\rm r}}))\right)}\, ,\label{eq:orientcorr}
\end{align}
where, $\langle\, \dotsb \rangle_{T}$ is the time average, quantifies the temporal evolution of the active noise term. The periodic cosine term results from the external torque and captures the induced circular motion. The rotational noise, quantified by $D_{\rm r}$, decorrelates the orientation on long-time scales. This decorrelation is described by the exponential term in Eq.~\eqref{eq:orientcorr}. The double exponential reflects the additional orientation correlation on short time scales imposed by the inertial damping rate $\tau_{\rm r}^{-1}$. Consequently, the particle dynamics non-trivially depend on the orientation, even in the short- and long-time limits. In the short-time limit the MSD is given by
\sj{
\begin{equation}
\label{eq:msdshort} 
\langle({\bf R(t)}-{\bf R_0})^2\rangle = \langle \dot{\bf R}^2 \rangle t^2 
\end{equation}
with
\begin{align}
\langle \dot{\bf R}^2 \rangle = 2D/\tau+\mathfrak{f}(\mathfrak{D}_0,\mathfrak{D}_1,\mathfrak{D}_2)V_{p}^{2} \quad . \label{eq:R2}
\end{align}
The first term is the equilibrium solution for a passive particle, and the second term arises from the active motion term. The latter is proportional to $V_p^2$, i.e.\ the kinetic energy injected by the propulsion. This contribution is quantified by the ratio of competing time scales, i.e.\ the dimensionless \textit{delay numbers}
\begin{equation}
\delaynum=D_{\rm r} \tau_r, \quad \delaynumw=\omega \tau_{\rm r}, \quad \delaynumx=\tau_{\rm r}/\tau \quad ,
\end{equation}
through the function 
\begin{align}
\mathfrak{f}(\mathfrak{D}_{0},\mathfrak{D}_{1},\mathfrak{D}_{2}) = \mathfrak{D}_{2} e^{\mathfrak{D}_{0}}&\operatorname{Re}\big[\mathfrak{D}_{0}^{-(\mathfrak{D}_{0}-\boldsymbol{i}\mathfrak{D}_{1}+\mathfrak{D}_{2})}\nonumber\\&\times\Gammainc(\mathfrak{D}_{0}-\boldsymbol{i}\mathfrak{D}_{1}+\mathfrak{D}_{2},\mathfrak{D}_{0})\big]\quad , 
\end{align}
where $\operatorname{Re}$ denotes the real part and $\Gammainc$ is the lower incomplete gamma function.}  
The long-time behaviour of the motion is diffusive, with the long-time diffusion coefficient 
\sj{
\begin{align}
D_\text{L} = & D+\frac{V_{p}^{2}}{2}\mathfrak{t}(\tau_{\rm r},\delaynum,\delaynumw) \quad . \label{eq:msdlong}
\end{align} }
In Eq.~\eqref{eq:msdlong}, the first term is the passive diffusion coefficient and the second term represents the contribution from the driving force with persistence time given by
\begin{align}
\label{eq:perstime}
\mathfrak{t}(\tau_{\rm r},\delaynum,\delaynumw) = \tau_r e^{\delaynum}\operatorname{Re}\left[\delaynum^{-(\delaynum-\boldsymbol{i}\delaynumw)}\Gammainc(\delaynum-\boldsymbol{i}\delaynumw,\delaynum)\right] \quad .
\end{align}
Equation~\eqref{eq:msdlong} is similar to the active Brownian motion model, where the persistence time $1/D_{\rm r}$ is replaced by Eq.~\eqref{eq:perstime}. The long-time diffusion coefficient is therefore a function of the inertial correlations introduced by $J$ through $\delaynum$. This starkly contrasts with passive Brownian motion, which assumes an inertia-independent diffusion coefficient.

\subsection{Parameter extraction}

\begin{figure*}[tb]
	\includegraphics[width=\textwidth]{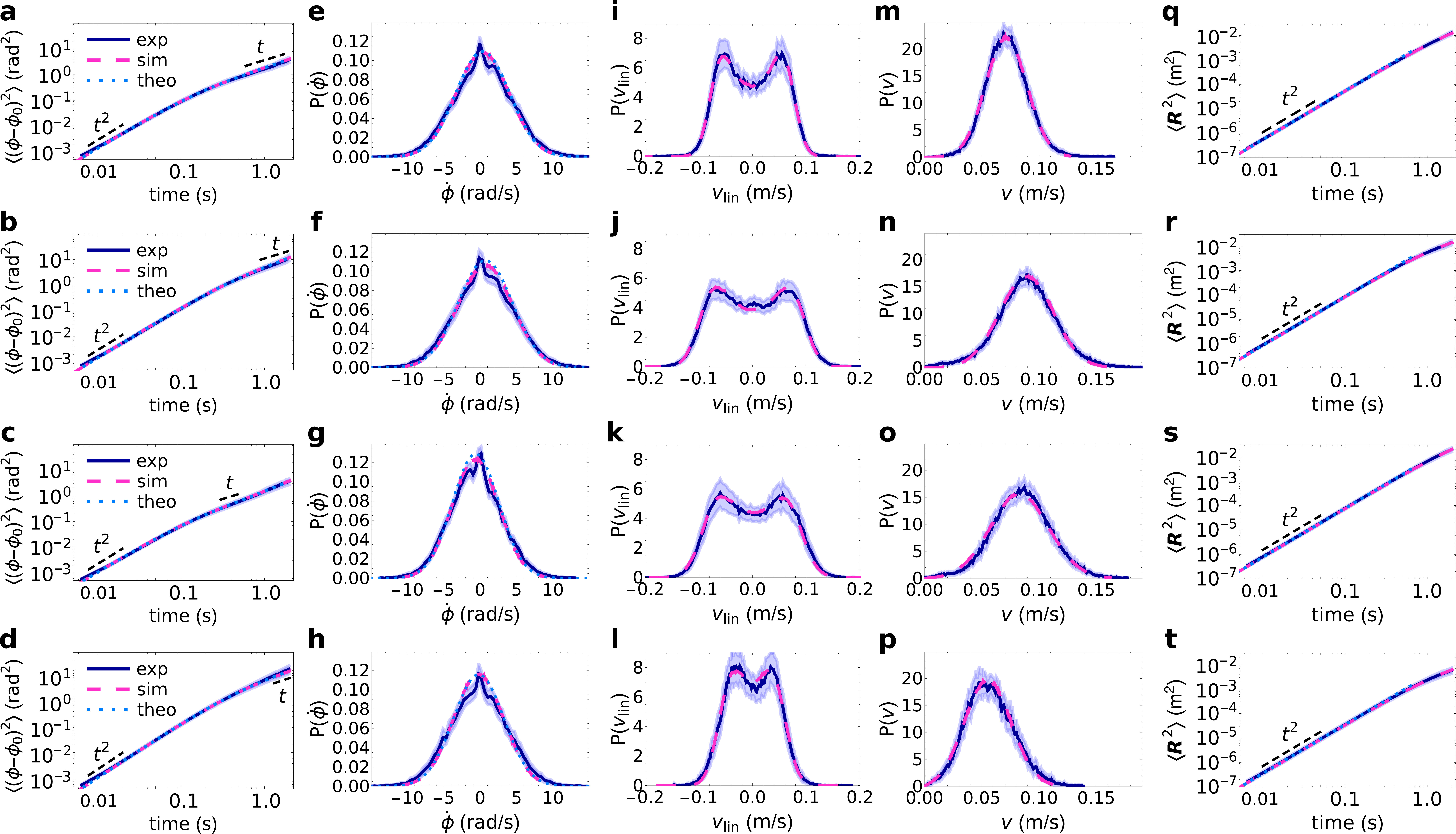}
	\caption{\label{Fig:Compare} \textsf{\textbf{Determination of model parameters for the generic, carrier, tug and ring particles (top to bottom).} \textbf{a}-\textbf{d} Rotational mean squared displacement, \textbf{e}-\textbf{h} Rotational velocity distribution, \textbf{i}-\textbf{l} Linear velocity distribution, \textbf{m}-\textbf{p} Absolute velocity distribution, \textbf{q}-\textbf{t} Translational mean squared displacement. \sj{Solid dark blue and ashed magenta curves show the expeirmental data and simulation results, respectively. Dotted light blue plots are the theoretical solutions.} The parameter values listed in Supplementary Table 1.}}
\end{figure*}

Equations \eqref{eq:R2} and \eqref{eq:msdlong} depend non-trivially on six independent parameters. They are determined by fitting the MSD given by Eq.~\eqref{eq:msdshort} and the linear and absolute velocity distributions, obtained by numerically solving Eqs.~\eqref{eq:lgm1} and \eqref{eq:lgm2}, to the measurements. The measurements and fitting curves for the four different particle types are summarized in Fig.~\ref{Fig:Compare}. The angular MSDs in Fig.~\ref{Fig:Compare}\textbf{a}-\textbf{d} show a ballistic short-time regime and a diffusive long-time regime (dashed lines) from which we can determine $\tau_{\rm r}$ and $D_{\rm r}$, respectively. 
The $\dot{\phi}$-distribution in Fig.~\ref{Fig:Compare}\textbf{e}-\textbf{h} is a shifted Gaussian. The minor deviations at small velocities are caused by the finite tracking accuracy. The first moment of this distribution gives the mean angular velocity $\omega$. The parameters $\tau$, $D$ and $V_p$ are extracted from the linear and absolute velocity distributions (Fig.~\ref{Fig:Compare}\textbf{i}-\textbf{p}) and the translational MSDs (Fig.~\ref{Fig:Compare}\textbf{q}-\textbf{t}), which can be directly fitted by Eq.~\eqref{eq:R2}. The linear velocity distribution is not a simple Gaussian, but shows a double peak related to the activity. The absolute velocity distribution also clearly deviates from the two-dimensional Maxwell-Boltzmann distribution of passive particles, especially, the maximum is shifted by the propulsion force. The translational MSD mainly depicts the ballistic short-time behaviour, because the persistence length of our particles is of the order of the system size.  
\begin{figure*}[tb]
	\includegraphics[width=\textwidth]{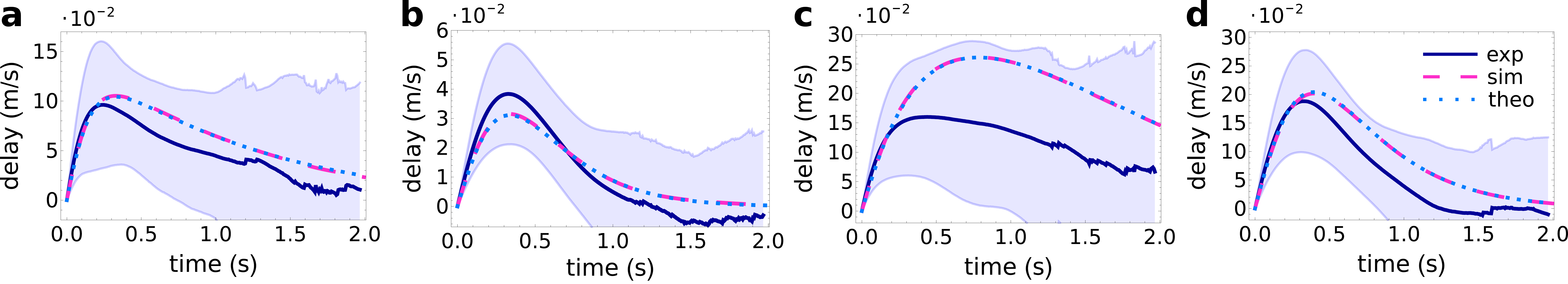}
	\caption{\label{Fig:DelayF} \textsf{\textbf{Time-dependence of the delay function.} Delay functions for the \textbf{a} generic, \textbf{b} carrier, \textbf{c} tug and \textbf{d} ring particle. \sj{The solid dark blue, dashed magenta and dotted light blue curve plot experimental, simulated and theoretical results, respectively.} Experimental uncertainties are expressed as the standard deviation. \sj{The parameters are those used in Fig.~\ref{Fig:Compare}.}}}
\end{figure*}
To test the parameters on an independent quantity, we systematically compared the model with the measured inertial delay. We define the correlation function
\sj{
\begin{align}
&	C(\dot{\bf R}(t),{\bf n}(t))=\langle \dot{\bf R}(t) \cdot {\bf n}(0) \rangle_T - \langle \dot{\bf R}(0) \cdot {\bf n}(t)  \rangle_T \quad  , \label{eq:delay}
\end{align} }
i.e.\ the average difference between the projection of the orientation on the initial velocity and projection of the velocity on the initial orientation. \sj{This function starts at zero and re-approaches zero in the limit $t\rightarrow\infty$.} In overdamped systems, Eq.\eqref{eq:delay} is zero at all times. In the underdamped case the velocity direction pursues the orientation and $C(\dot{\bf R}(t),{\bf n}(t))$ reaches its maximum after a specific delay. Pronounced peaks, related to the decay numbers and $\tau_{\rm r}$ are observed in Fig.~\ref{Fig:DelayF}\textbf{a}-\textbf{d}. The measurements, simulation results and analytical expressions using the parameters determined from Fig.~\ref{Fig:Compare} are consistent within the experimental accuracy.

\subsection{Inertial dependence}

\begin{figure}[tb]
	\includegraphics[width=\columnwidth]{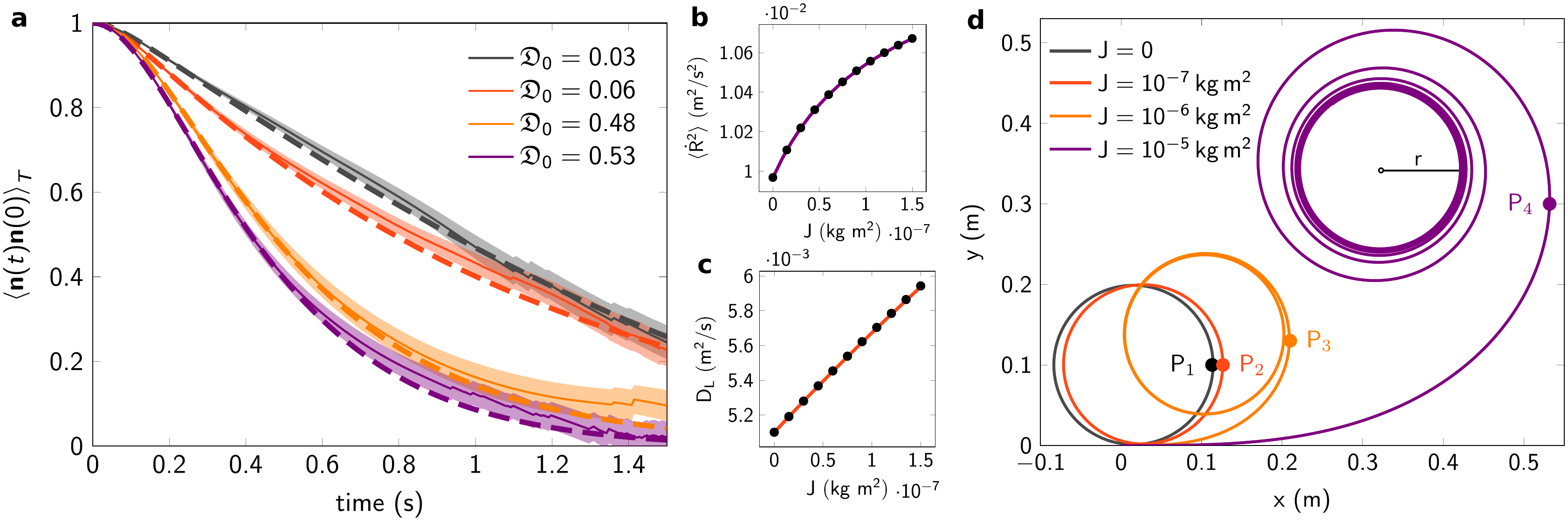}
	\caption{\label{Fig:MI} \textsf{\textbf{Particle dynamics dependence on rotational inertial delay.} 
	\textbf{a} Time dependence of orientational correlation functions. Solid lines represent the measurements, error bands represent the standard error of the mean. Dashed lines are the analytic results using the parameters from Fig.~\ref{Fig:Compare}. \textbf{b} Slope of the ballistic regime, i.e.\ the second moment of the velocity $\mathsf{\langle \dot{\bf R}^2 \rangle}$, as a function of $\mathsf{J}$. (the circles and solid line are the numerical results and the analytic solution to Eq.~\eqref{eq:R2}, respectively). The model parameters (except $\mathsf{J}$) are those used in Fig.~\ref{Fig:MSD}. \textbf{c} Long-time diffusion coefficient $\mathsf{D_\text{L}}$ as a function of $\mathsf{J}$ (the circles and solid line are the numerical results and the analytic solution to Eq.~\eqref{eq:msdlong}, respectively). \textbf{d} Trajectories of active particles under a constant torque applied at $\mathsf{t_0}$. As $\mathsf{J}$ increases the turn-around manoeuvre becomes increasingly difficult, so the distance and time increase until the turning point $\mathsf{P}_{1,2,3,4}$ is reached.}}
\end{figure}

Strikingly, both the short- and long-time particle dynamics in our system depend on the delay number $\delaynum$. The fundamental reason is the additional orientational correlation in Eq.~\eqref{eq:orientcorr}, which is delayed by the rotational friction rate $\tau_{\rm r}^{-1}$. The exponent in this expression represents the MSD of $\phi$, which is dominated by order $t^2$ at short times and order $t$ at long times. Consequently, neglecting external torque, this function follows a Gaussian decay at short times and an exponential decay at long times. The significance of the inertial delay is quantified by $\delaynum$. For small $\delaynum$, the correlation approaches the overdamped result and for large $\delaynum$ the correlation time is significantly delayed by $\tau_{\rm r}$. To confirm this prediction, we compare the measured correlation functions and the solutions of Eq.~\eqref{eq:orientcorr}. The results are consistent, as shown in Fig.~\ref{Fig:MI}\textbf{a}.\\
The numerical and analytical dependence of the ballistic and diffusive regimes on the moment of inertia are displayed in Fig.~\ref{Fig:MI}\textbf{b},\textbf{c}, which show that $\langle \dot{\bf R}^2 \rangle$ and $D_\text{L}$ increase with $J$.
The effects of finite $J$ can be simply demonstrated mathematically by expanding Eqs.~\eqref{eq:R2} and \eqref{eq:msdlong} in the limit $J\rightarrow0,\infty$. As $J$ vanishes, we find that
\sj{
\begin{equation}
\lim_{J\rightarrow0}\langle({\bf R(t)}-{\bf R_0})^2\rangle = \left(2D\frac{\xi}{M} + V_{p}^{2}\frac{\xi}{\xi+M D_{\rm r}} \right) t^2 ,
\end{equation} }
which agrees with results reported in \cite{Mokhtari2017}. For infinitely large $J$ we obtain
\sj{
\begin{equation}
\lim_{J\rightarrow\infty}\langle({\bf R(t)}-{\bf R_0})^2\rangle = \left(2D\frac{\xi}{M} + V_{p}^{2}\right) t^2 \quad ,
\end{equation} }
which is simply the sum of the thermal and injected kinetic energies. 
For the long-time diffusion constant \sj{and small moments of inertia, the asymptotic behaviour is 
\begin{equation}
D_\text{L} = D+\frac{V_{p}^{2}}{2D_{\rm r}}+\frac{V_{p}^{2}}{2\xi_{\rm r}}J+ \mathcal{O}(J^2) \, ,
\end{equation} }
which intuitively demostrates, how, the leading order $J$ increases the persistence time (namely by a linear term proportional to $(\gammar)^{-1}$). The dependence of $D_\text{L}$ on $\delaynum$ has no upper bound, and its asymptotic behaviour is described by
\sj{
\begin{equation}
D_\text{L} =  D+V_{p}^{2}\sqrt{\frac{\pi}{8D_{\rm r}\xi_{\rm r}}}\sqrt{J}+ \mathcal{O}\left(\sqrt{J}^{\,\,-1}\right) \, .
\end{equation} }
The origin of this dependence can be intuitively understood by considering the turn-around manoeuvre of a simple noise-free active particle. When a torque is applied perpendicularly to the velocity, the particle will turn around at point $P$ and eventually approach circular motion. As the moment of inertia quantifies the resistance of a particle to changing its angular momentum, a particle with low $J$ will turn faster than one with high $J$, as shown in Fig.~\ref{Fig:MI}\textbf{d}. This applies only to the transient states, where $\ddot{\phi}\neq0$. In the steady state, the radius $r$ of the final circle is independent of $J$. The angular momentum of an active particle with random reorientations is constantly changing. Its inertia resists these changes and modifies the distribution of reorientations directly opposing the effect of rotational noise.

\section{Discussion}

\begin{figure}[tb]
	\includegraphics[width=0.65\columnwidth]{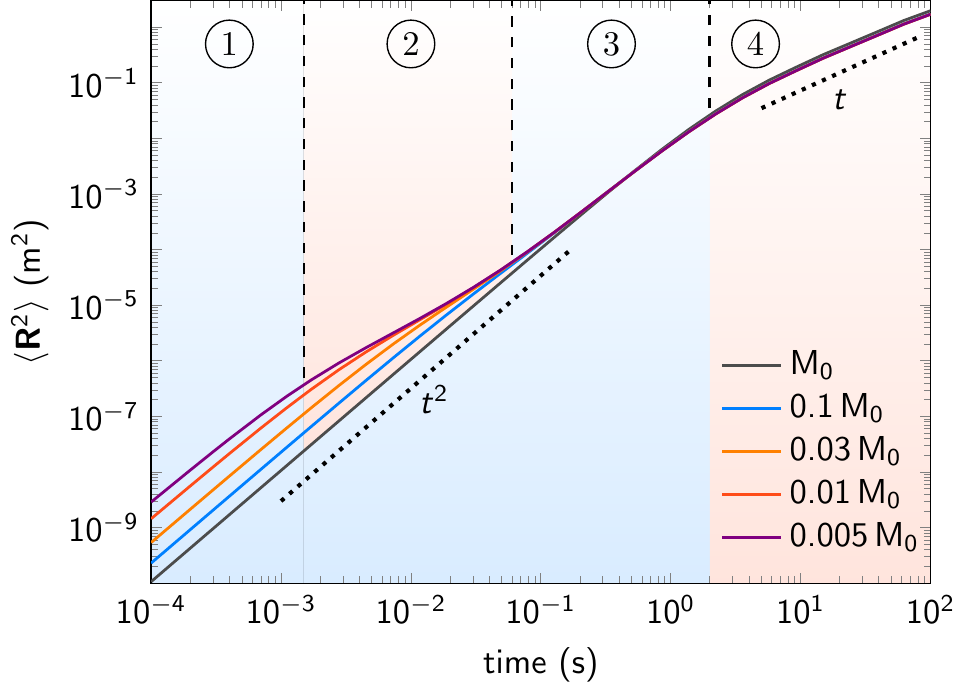}
	\caption{\label{Fig:MSD} \textsf{\textbf{Time dependence of MSD for gradually decreasing density.} 
	MSD of hypothetical particles with successively reduced density (i.e.\ reduced $\mathsf{M}$ and $\mathsf{J}$). The other parameters are fixed as $\mathsf{M_0=4\,\text{g}}$, $\mathsf{J_0=1.5\times10^{-7}\,\textsf{kg}\textsf{m}^2}$, $\mathsf{\gammat=6.5\,\textsf{s}^{-1}}$, $\mathsf{\gammar=5.5\,\textsf{s}^{-1}}$ $\mathsf{\difft=1\times10^{-4}\,\textsf{m}^2/\textsf{s}}$, $\mathsf{D_{\rm r}=1\,\textsf{s}^{-1}}$, $\mathsf{V_p=0.1\,\textsf{m}/\textsf{s}}$, $\mathsf{\omega=0}$. When $\textsf{M}=\textsf{M}_0$, two regimes are visible. When the mass drops below $\mathsf{0.01\,\textsf{M}_0}$ the MSD divides into four regimes: 1.~inertial ballistic, 2.~short-time diffusive, 3.~active ballistic and 4.~active diffusive. In the limit of large damping and vanishing torque the 1--2, 2--3 and 3--4 transition times are given by $\mathsf{M/\xi}$, $\mathsf{D/2V_p^2}$ and $\mathsf{D_{\rm r}^{-1}}$, respectively.}}
\end{figure}

Our observations demonstrate the profound influence of inertia on the long and short-time dynamics of self-propelled particles. 
Considering the relevance of inertia\cite{Jaeger1996}, our model is applicable to various systems, such as levitating\cite{Workamp2018,Farhadi2018} and floating\cite{Harth2018} granular particles and dusty plasmas\cite{Ivlev2015}. 
Our model predicts that microswimmers perform a short-time ballistic motion like passive particles, but in practice, their motion also depends on their specific propulsion mechanism\cite{Maass2016,Michelin2017} and hydrodynamic effects\cite{Pesce2014,Elgeti2015}.
Generally, the inertial effects will depend on the corresponding time scales in the system. In numerical experiments, this can be demonstrated by gradually reducing the density of hypothetical particles, retaining all other parameters as constants. At very low densities, the MSD exhibits four different regimes: short-time ballistic, short-time diffusive, active ballistic and long-time diffusive regime (see Fig.~\ref{Fig:MSD}).\\
The long-time diffusion coefficient of passive particles is independent of inertia and is related to the friction coefficient via the Stokes-Einstein relation. However, for actively moving particles we find an explicit dependence on the moment of inertia (with no explicit dependence on the total mass $M$). This finding illustrates the importance of $J$ for macroscopic self-propelled particles. While mass distribution and shape are generally important for efficient motion of animals\cite{kram1997three,raichlen2006effects,dudley1992power,dudley2002mechanisms} and adaption to the environment\cite{xu2015,Lin2018}, our results suggests that $J$ can be exploited in novel control strategies for active matter. Biological organisms cannot rapidly vary their mass, but they can change $J$ by moving their limbs. For instance, cheetahs use tail motion to stabilize fast turns\cite{Wilson2013}. By decreasing $J$, animals can more easily dodge obstacles or predators; conversely, they can increase $J$ to faster explore a large area. Even under conditions, where animals cannot control their rotational deflections, such as aerodynamic turbulence, or during random collisions with neighbours\cite{Zuriguel2014}, they could control their movements through variations of $J$. 

\section{Data availability}
The data that support the plots within this paper and other findings of this study are available from the corresponding authors upon request.

\section{Methods}

\textbf{Particle Fabrication.}
Four particle types were designed and printed:
The \textit{generic particle} consists of a cylindrical core (diameter 9~mm, length 4~mm) topped by a cylindrical cap (diameter 15~mm, length 2~mm). Beneath the cap, seven tilted cylindrical legs (each of diameter 0.8~mm) were attached in parallel in a regular heptagon around the core. The legs lift the bottom of the body by 1 mm above the surface. The typical mass was about $m=0.76\,\text{g}$. From the mass and shape of the particle the moment of inertia was approximated as $J=1.64\times10^{-8}\,\text{kg}~\text{m}^2$. To vary the propulsion velocity of the particles, we printed five types with different leg inclination angles 0, 2, 4, 6 and 8 degrees.\\
The \textit{carrier particle} was fabricated with the same core as \textit{generic}, but its cap was topped with a 1~mm tall, 8.5~mm diameter cylinder. The carrier socket held two galvanised steel washers, each with an outer diameter of 16~mm and a mass of 1.6~g. The leg inclination of \textit{carrier particles} was fixed at 2~degrees, and mass and moment of inertia were $m=4.07\,\text{g}$, $J=1.46\times10^{-7}\,\text{kg}~\text{m}^2$, respectively.\\
The \textit{tug particle} was a \textit{generic} with a fixed leg inclination of 2~degree and thinner core (diameter 4~mm). This core held a hexagonal M5 threaded galvanised steel nut with a short diagonal and height of 8~mm and 3.75~mm, respectively. The mass and moment of inertia were $m=1.57\,\text{g}$ and $J=2.54\times10^{-8}\,\text{kg}~\text{m}^2$, respectively.\\
The \textit{ring particle} had a leg inclination of 4~degree and a ring shaped cap with a hole (diameter 9~mm) in the middle. The mass and moment of inertia were $m=0.33\,\text{g}$ and $J=1.26\times10^{-8}\,\text{kg}~\text{m}^2$, respectively.\\
All particles were labelled with a simple high contrast image allowing the detection software to identify the particle's position and orientation. The particles were printed from a proprietary methacrylate based photopolymer (FormLabs Grey V3, FLGPGR03) of typical density 1.1(1)g/L at a precision of 0.05~mm. They were subsequently cleaned in high purity ($>97\%$) isopropyl alcohol in a still bath, followed by an ultrasound bath, then hardened by three 10-min bursts under four 9~W UVA bulbs. Finally, irregularities were manually filed away and the label sticker was attached.\\

\textbf{Experimental setup.}
The vibrobots were excited by vertical vibrations generated by a circular acrylic baseplate (diameter 300~mm, thickness 15~mm) attached to an electromagnetic shaker (Tira TV 51140) and surrounded by a barrier to confine the particles. The tilt of the plate was adjusted with an accuracy of $10^{-2}$~degrees. The vibration frequency and amplitude was set to $f=80\,\mathrm{Hz}$ and $A = 66(4)\mathrm{\mu m}$, respectively, guaranteeing stable excitation with peak accelerations of $1.7(1)\,g$ (measured by four LIS3DH accelerometers). Experiments were recorded using a high-speed camera system (Allied Vision Mako-U130B) operating at up to 152 fps with a spatial resolution of $1024 \times 1024$ pixels. Single particles were tracked to sub-pixel accuracy using standard image recognition methods. Multiple single trajectories were recorded for each particle, until 10 min of data were acquired. Events involving particle-border collisions were discarded.\\

\sj{\textbf{Analytic results.}
The rotational behaviour of the particle was obtained by stochastic integration\cite{risken1996fokker} of Eq.\,\eqref{eq:lgm2}. The angular frequency and angular coordinate we obtained as
\begin{alignat}{1}
\dot{\phi}(t) &=\omega +(\dot{\varphi}_{0}-\omega)e^{-\xi_{r} t/J}\nonumber\\&+\sqrt{2D_{r}}\,\frac{\xi_{r}}{J}e^{-\xi_{r} t/J}\int_{0}^{t^{}}dt^{'}e^{\xi_{r} t^{'}/J}\tau_{\rm st}(t^{'})\, ,
\label{AngularMotion}
\end{alignat}
and
\begin{alignat}{1}
\phi(t) &=\varphi_{0}+\omega t+\frac{\omega-\dot{\varphi}_{0}}{\xi_{r}}J\left(e^{-\xi_{r} t/J}-1\right)+\sqrt{2D_{r}}\,\frac{\xi_{r}}{J}\nonumber\\&\times\int_{0}^{t}dt^{'}e^{-\xi_{r} t^{'}/J}\int_{0}^{t^{'}}dt^{''}e^{\xi_{r} t^{''}/J}\tau_{\rm st}(t^{''})\, ,
\label{phiIntegralForm}
\end{alignat}
respectively. Here, $\phi_{0}$ and $\dot{\varphi}_{0}$ are initial angle and angular velocity, respectively, and the initial time was set to zero. As $\dot{\phi}(t)$ and $\phi(t)$ are both linear combinations of Gaussian variables, the corresponding probability distributions are also Gaussian. Thus, by calculating the mean
\begin{alignat}{1}
\langle \phi(t) \rangle &=\varphi_{0}+\omega t+\frac{\omega-\dot{\varphi}_{0}}{\xi_r}J\left(e^{-\xi_{\rm r} t/J}-1\right)\, ,
\label{meanphi}
\end{alignat}
and the variance
\begin{alignat}{1}
\mu(t)= 2D_{\rm r}t+\frac{2D_{\rm r}}{\xi_{\rm r}}J\left( e^{-\xi_{\rm r} t/J}-1-\frac{\left(e^{-\xi_{\rm r} t/J}-1\right)^{2}}{2}\right)\, ,
\label{variancephi}
\end{alignat}
one obtains the angular probability distribution
\begin{alignat}{1}
P(\phi,t) &=\frac{1}{\sqrt{2\pi \mu(t) }}\exp\left(\frac{-\left(\phi-\langle \phi(t) \rangle\right)^{2}}{2 \mu(t)}\right)\, .
\label{phiDist}
\end{alignat}
At times much longer than the reorientation time scale $1/D_{\rm r}$ and the rotational friction rate $J/\xi_{\rm r}$, the variance of the angular distribution far exceeds $2\pi$, while the mean cycles between $0$ and $2\pi$. This behaviour converges to the stationary state with a uniform distribution of $\phi$. At times much longer than the rotational friction rate $J/\xi_{\rm r}$, the stationary distribution of the angular velocity reduces to
\begin{alignat}{1}
P(\dot{\phi}) &=\sqrt{\frac{J}{2\pi D_{\rm r}\xi_{\rm r}}}\exp\left(\frac{-J(\dot{\phi}-\omega)^{2}}{2 D_{\rm r}\xi_{\rm r}}\right)\, .
\label{phiDotDist}
\end{alignat}
The width of this distribution is inversely proportional to the moment of inertia.

From the translational equation of motion i.e.\ Eq.\,\eqref{eq:lgm1}, the velocity in the laboratory frame of reference is obtained as
\begin{alignat}{1}
\boldsymbol{\dot{R}}(t) &= \boldsymbol{\dot{R}}_{0}e^{-\xi t/M}+\frac{\xi}{M} V_{p}e^{-\xi t/M}\int_{0}^{t}dt^{'}e^{\xi t^{'}/M} \,{\bf n}(t^{'})\nonumber\\&+\sqrt{2D}\,\frac{\xi}{M}e^{-\xi t/M}\int_{0}^{t}dt^{'}e^{\xi t^{'}/M}\,{\bf f}_{\rm st}(t^{'})\, ,
\label{velx_lab}\, ,
\end{alignat}
where the initial velocity is denoted by $\boldsymbol{\dot{R}}_{0}$. The centre-of-mass position of a particle beginning its motion from the origin is calculated as
\begin{alignat}{1}
\boldsymbol{R}(t) & =\boldsymbol{R}_{0} + \boldsymbol{\dot{R}}_{0}\frac{M}{\xi}\big(1-e^{-\xi t/M}\big)+\frac{\xi}{M} V_{p}\int_{0}^{t}dt^{'}e^{-\xi t^{'}/M}\nonumber\\&\times\int_{0}^{t^{'}}dt^{''}e^{\xi t^{''}/M} {\bf n}(t^{''})+\sqrt{2D}\,\frac{\xi}{M}\int_{0}^{t}dt^{'}e^{-\xi t^{'}/M}\nonumber\\&\times\int_{0}^{t^{'}}dt^{''}e^{\xi t^{''}/M}\,{\bf f}_{\rm st}(t^{''})\, ,
\label{x_lab}
\end{alignat}
The mean square displacement $\langle \boldsymbol{R}^{2} \rangle$ is obtained in the following integral form
\begin{alignat}{1}
&\langle\boldsymbol{R}^{2}(t)\rangle  = \boldsymbol{\dot{R}}_{0}^{2}\frac{M^{2}}{\xi^{2}}\big(1-e^{-\xi t/M}\big)^{2}+2 V_{p}\big(1-e^{-\xi t/M}\big)\nonumber\\&\times\int_{0}^{t}dt^{'}e^{-\xi t^{'}/M}\int_{0}^{t^{'}}dt^{''}e^{\xi t^{''}/M} \,\boldsymbol{\dot{R}}_{0}.\langle{\bf n}(t^{''})\rangle+\frac{\xi^{2}}{M^{2}} V_{p}^{2}\nonumber\\&\times\int_{0}^{t}dt^{'}e^{-\xi t^{'}/M}\int_{0}^{t^{'}}dt^{''}e^{\xi t^{''}/M}\int_{0}^{t}d\tau^{'}e^{-\xi \tau^{'}/M}\nonumber\\&\times\int_{0}^{\tau^{'}}d\tau^{''}e^{\xi \tau^{''}/M} \,\langle{\bf n}(t^{''}).{\bf n}(\tau^{''})\rangle+4Dt\nonumber\\&+\frac{4D}{\xi}M\Big( e^{-\xi t/M}-1-\frac{1}{2}(e^{-\xi t/M}-1)^{2}\Big),
\label{xsqPysq_lab}
\end{alignat}
where $\langle{\bf n}(t)\rangle =e^{-\mu(t)/2}\big(\cos \langle\phi(t)\rangle\, ,\,\sin \langle\phi(t)\rangle\big)$ and $\langle{\bf n}(t_{1}).{\bf n}(t_{2})\rangle$ is defined by
\begin{align}
& \langle{\bf n}(t_{1}).{\bf n}(t_{2})\rangle=e^{-D_{r}|t_{1}-t_{2}|}e^{D_{r}J/\xi_{r}}\exp\Bigg[\frac{-D_{r}}{\xi_{r}}J\nonumber\\&\times\Big( e^{-\frac{\xi_{r}}{J} |t_{1}-t_{2}|}+e^{-\frac{\xi_{r}}{J} (t_{1}+t_{2})}-\frac{1}{2}(e^{-2\frac{\xi_{r}}{J} t_{1}}+e^{-2\frac{\xi_{r}}{J} t_{2}})\Big)\Bigg]\nonumber\\&\times\cos\Big[\omega(t_{1}-t_{2})+\frac{\omega-\dot{\phi}_{0}}{\xi_{r}}J(e^{-\frac{\xi_{r}}{J} t_{1}}-e^{-\frac{\xi_{r}}{J} t_{2}})\Big].
\label{corr_cos_cos}
\end{align}
}
\sj{
The inertial delay correlation function Eq.~\eqref{eq:delay} is given by
\begin{align}
&	\langle \dot{\bf R}(t) \cdot {\bf n}(0) \rangle_T - \langle \dot{\bf R}(0) \cdot {\bf n}(t)  \rangle_T = \nonumber\\
&\quad V_{p}\delaynumx e^{\delaynum}\delaynum^{(\delaynumx-\delaynum)}e^{-t/\tau}\nonumber\\
&\times\operatorname{Re}\Bigg[\delaynum^{\boldsymbol{i}\delaynumw}\Big(\delaynum^{-2\delaynumx}\Gammainc(\delaynum-\boldsymbol{i}\delaynumw+\delaynumx,\delaynum)\nonumber\\
&\quad\qquad-e^{2t/\tau}\delaynum^{-2\delaynumx}\Gammainc(\delaynum-\boldsymbol{i}\delaynumw+\delaynumx,\delaynum e^{-t/\tau_{\rm r}})\nonumber\\
&\quad\qquad-\Gammainc(\delaynum-\boldsymbol{i}\delaynumw-\delaynumx,\delaynum e^{-t/\tau_{\rm r}})\nonumber\\
&\quad\qquad+\Gammainc(\delaynum-\boldsymbol{i}\delaynumw-\delaynumx,\delaynum)\Big)\Bigg]
\quad  , \label{eq:delay_method}
\end{align} }

%

\section{Acknowledgements} 
We acknowledge funding by the German Research Foundation (Grant No. SCHO 1700/1-1 and LO 418/23-1).

\section{Author contributions}
C.S.\ designed the experimental setup.
C.S.\ and A.L.\ carried out the experiments.
C.S.\, S.J.\ and A.L.\ analyzed the measurements.
S.J.\ and C.S.\ wrote the simulation code.
S.J.\ and C.S.\ performed and analyzed the simulations.
S.J.\ developed the theoretical results.
All authors discussed the results and wrote the manuscript.

\section{Additional information}

\textbf{Supplementary information} accompanies this paper 

\textbf{Competing financial interests:}
The authors declare no competing financial interests.

\end{document}